\documentclass[twocolumn,aps,amsmath,amssymb,superscriptaddress]{revtex4-2}

\usepackage{newtxtext}
\usepackage{newtxmath}
\usepackage{microtype}

\usepackage{graphicx}
\usepackage{dcolumn}
\usepackage{bm}
\usepackage{multirow}
\usepackage{amssymb}
\usepackage{graphicx}
\usepackage{xcolor}
\definecolor{TighnariBrown}{RGB}{141,87,41}
\definecolor{TighnariGreen}{RGB}{40,114,70}
\definecolor{TighnariYellow}{RGB}{247,191,99}
\definecolor{PRLBlue}{RGB}{46,48,146}
\usepackage[
    colorlinks,
    citecolor=PRLBlue,
    linkcolor=PRLBlue,
    urlcolor=PRLBlue,
]{hyperref}
\usepackage{indentfirst}
\usepackage{cases}
\usepackage{braket}
\usepackage{booktabs}

\usepackage{orcidlink}
\newcommand{\ror}[2]{\href{https://ror.org/#1}{#2}}

\def \i{\mathrm{i}}
\def \q{\bm{q}}
\def \kuohao#1{\left(#1\right)}

\newcommand{\expup}[1]{\mathrm{e}^{#1}}

\def \mat#1{\begin{pmatrix}#1\end{pmatrix}}

\makeatletter
\let\oldaddcontentsline\addcontentsline
\renewcommand{\addcontentsline}[3]{}
\makeatother

\begin{document}
\title{Nonequilibrium Dynamics of Dirac Quantum Criticality in Imaginary Time}

\author{Yin-Kai Yu\,\orcidlink{0009-0006-2589-2772}}
\affiliation{Guangdong Provincial Key Laboratory of Magnetoelectric Physics and Devices,\\ \ror{0064kty71}{Sun Yat-Sen University}, Guangzhou 510275, China}
\affiliation{School of physics, \ror{0064kty71}{Sun Yat-Sen University}, Guangzhou 510275, China}
\affiliation{Beijing National Laboratory for Condensed Matter Physics\\ \& \ror{05cvf7v30}{Institute of Physics, Chinese Academy of Sciences}, Beijing 100190, China}
\affiliation{\ror{05qbk4x57}{University of Chinese Academy of Sciences}, Beijing 100049, China}

\author{Zhi Zeng\,\orcidlink{https://orcid.org/0009-0002-3471-6405}}
\affiliation{Guangdong Provincial Key Laboratory of Magnetoelectric Physics and Devices,\\ \ror{0064kty71}{Sun Yat-Sen University}, Guangzhou 510275, China}
\affiliation{School of physics, \ror{0064kty71}{Sun Yat-Sen University}, Guangzhou 510275, China}

\author{Yu-Rong Shu\,\orcidlink{https://orcid.org/0000-0001-7277-8346}}
\affiliation{School of Physics and Materials Science, \ror{05ar8rn06}{Guangzhou University}, Guangzhou 510006, China}

\author{Zi-Xiang Li\,\orcidlink{0000-0002-3941-365X}}
\email{zixiangli@iphy.ac.cn}
\affiliation{Beijing National Laboratory for Condensed Matter Physics\\ \& \ror{05cvf7v30}{Institute of Physics, Chinese Academy of Sciences}, Beijing 100190, China}
\affiliation{\ror{05qbk4x57}{University of Chinese Academy of Sciences}, Beijing 100049, China}

\author{Shuai Yin\,\orcidlink{https://orcid.org/0000-0001-8534-9364}}
\email{yinsh6@mail.sysu.edu.cn}
\affiliation{Guangdong Provincial Key Laboratory of Magnetoelectric Physics and Devices,\\ \ror{0064kty71}{Sun Yat-Sen University}, Guangzhou 510275, China}
\affiliation{School of physics, \ror{0064kty71}{Sun Yat-Sen University}, Guangzhou 510275, China}

\date{\today}
\begin{abstract}
Quantum criticality within Dirac fermions harbors a plethora of exotic phenomena, attracting sustained attention in the past decades. Here, we explore the imaginary-time relaxation dynamics in a typical Dirac quantum criticality belonging to chiral Heisenberg universality class. Performing large-scale quantum Monte Carlo simulation, we unveil rich nonequilibrium critical phenomena from different initial states. In particular, we identify a non-stationary initial slip evolution characterized by an unconventional negative critical exponent $\theta=-0.84(4)$, corroborating the significant impact of fermionic critical fluctuations. Furthermore, we generalize the nonequilibrium scaling theory to incorporate both fermionic and bosonic critical modes, capturing their distinct relaxation behaviors. Armed with the scaling theory, we establish a new framework to investigate fermionic quantum criticality based on short-time dynamics, paving a promising avenue to fathoming quantum criticality in diverse fermionic systems with high efficiency.
\end{abstract}
\maketitle

\pdfbookmark[1]{Introduction}{sec1}
\textcolor{black}{\it Introduction}--- Quantum phase transitions, describing abrupt changes in ground states of quantum systems, are central topics in modern physics~\cite{Sachdevbook}. A prominent example is the interaction-driven quantum criticality in Dirac systems. Such transitions were originally discussed in high-energy physics to mimic chiral symmetry breaking and spontaneous mass generation~\cite{Gross1974prd}. Recently, owing to the inspiring experimental advances in graphene~\cite{Geim2009rmp} and topological materials~\cite{KaneReview,SCZhangReview}, quantum criticality in Dirac fermions has garnered increasing interest in condensed matter physics. Vast efforts have been paid in this field, including sophisticated renormalization group analyses~\cite{Sherer2017prb,Sherer2017prd,Janssen2023prb,Janssen2014prb,Knorr2016prb,Gracey2016prd,Roy2016JHEP,Gies2015prd,Scherer2018prb,Joseph2021review,Roy2016PRB,Moon2018PRB}, conformal bootstrap~\cite{Poland2019rmp}, quantum Monte Carlo simulation~\cite{Lang2019PRL,Liuzh2023prl,Li2018ScienceAdvances,Vishvanath2017NP,Wu2016PRB,Vaezi2022PRL,Sorella2018PRB,Xu2021PRL,Meng2020PRB,Chandrasekharan2013PRD,Guo2022PRB,Meng2023arXiv,Wang2023PRR,Li2024PRL,Li2024PRL2} and tensor network method~\cite{Corboz2018prx,Andreas2018prx}, resulting in tremendous achievements. It was shown that fluctuations from gapless Dirac fermions enormously fertilize the fundamental research of quantum criticality, not only contributing to the Gross-Neveu fixed point~\cite{Roy2016JHEP,Janssen2023prb,Janssen2014prb,Knorr2016prb,Gracey2016prd,Gies2015prd,Scherer2018prb,Joseph2021review,Poland2019rmp,Lang2019PRL,Liuzh2023prl,Li2018ScienceAdvances,Vishvanath2017NP,Wu2016PRB,Vaezi2022PRL,Sorella2018PRB,Xu2021PRL,Meng2020PRB,Chandrasekharan2013PRD,Meng2023arXiv,Wang2023PRR,Guo2022PRB,Sorella1992,Herbut2006prl,Herbut2009prb,Herbut2013prx,Sorella2016prx,Sherer2017prb,Sherer2017prd,Sheng2014science,Seifert2020prl,Janssen2021prb,Corboz2018prx,Andreas2018prx,Qi2022PRB,Wang2014NJP,Li2015NJP,Guo2021PRB,Assaad2022PRL,Hohenadler2019PRL,Scalettar2019PRL,Roy2016PRB,Moon2018PRB,Herbut2023arXiv,Otsuka2020PRB,Li2024PRL,Li2024PRL2}, which is among the simplest examples of quantum critical points that do not exhibit classical analogs, but also yielding a profound mechanism for the Landau-forbidden quantum criticality~\cite{Yao2017nc,Li2020PRB,Scherer2018prb2,Scherer2017prb2,Youyz2018prx,Yin2020prb,You2023PRB}. 

On the other hand, universal critical phenomena are manifested not only in the long-time equilibrium states but also in short-time nonequilibrium processes~\cite{Hohenberg1977rmp,Dziarmaga2010review,Polkovnikov2011rmp,Rigol2016review,Mitra2018arcmp}. For instance, in classical systems, the relaxation dynamics shows a non-stationary initial slip evolution in the short-time stage and an additional critical initial slip exponent, which is usually positive, is required to describe this phenomenon~\cite{Janssen1989,Lizb1995prl,Albano2011iop}. 
Similar short-time scaling behaviors also arise in real-time critical dynamics in quantum systems~\cite{Polkovnikov2013prl,Schmalian2014prl,Schmalian2015prb,Chiocchetta2015prb,Marino201prl,Yin2019prl,Yin2021prb,Marino_2022}.
Particularly, in Dirac systems, the field theory predicted that fluctuations from the Dirac fermions can induce a negative critical initial slip exponent~\cite{Yin2019prl}.

Aside from the real-time dynamics, imaginary-time dynamics in quantum systems is also of great interest and significance. As a routine unbiased approach to identify the ground state, the imaginary-time evolution works extensively in numerical simulations, such as quantum Monte Carlo (QMC) and tensor networks. Near a quantum critical point (QCP), it was shown that the imaginary-time critical dynamics (ITCD) demonstrates colorful universal scaling behaviors~\cite{Polkovnikov2011prb,De_Grandi_2013,PolkovnikovSandvik2013prb,Yins2014prb}. So far, the ITCD has been studied in various quantum systems, providing an abundance of intriguing perspectives in the field of quantum criticality~\cite{Yins2014prb,Shu2017prb,Yins2014pre,Yin2022prl,Yin2022prb,Zuo2021prb,Sandvik2015prl,Sandvik2023nature}.

Moreover, the imaginary-time dynamics also finds its practical applications in fast-developing quantum programmable devices including quantum circuits simulating fermionic systems~\cite{Motta2020naturephyscis,Nishi2021njp,Pollmann2021prxq}, partly spurred by the recent availability of noisy intermediate-scale quantum hardware and Rydberg atomic systems,  to explore various exotic quantum phases~\cite{Satzinger2021science,Semeghini2021science,Zoller2023pnas,Zoller2023fermion-qudit}. As a win-win case, these quantum devices also provide platforms to experimentally realize ITCD and demonstrate its power in determining the critical properties with high efficiency and scalability, circumventing difficulties induced by critical slowing down and divergent entanglement in conventional method based on equilibrium scaling~\cite{Zhang2023}.

However, nonequilibrium dynamics in Dirac quantum criticality remains largely unexplored. This kind of quantum criticality is fundamentally different from the previously studied paradigms of ITCD~\cite{Yins2014prb}, which feature a single type of critical fluctuation such as bosonic order parameters. The key difference lies in the coexistence of fermionic and bosonic critical fluctuations in the Dirac criticality. Given the many intriguing features of gapless Dirac fermions in equilibrium, it is a compelling question how these fluctuations influence nonequilibrium dynamics.

Following this line of inquiry, we investigate the ITCD of a paradigmatic Dirac-fermion QCP, namely the chiral Heisenberg Gross-Neveu transition hosted by the honeycomb Hubbard model~\cite{Sorella1992,Herbut2006prl,Herbut2009prb,Herbut2013prx,Sorella2016prx}. Employing sign-problem-free QMC simulations~\cite{Sorella1989EPL,AssaadReview,Li2019Review,Troyer2005PRL,Wu2005PRB,Li2015PRB,Li2016PRL,Xiang2016PRL,Berg2012Science}, we find that the system exhibits rich dynamic scaling behaviors for different initial states. We generalize the nonequilibrium scaling theory by unifying the scaling of both fermionic and bosonic correlations.
In particular, we numerically find a negative critical initial slip exponent arising in the ITCD, in stark contrast to the bosonic systems, where the exponent is typically positive. We discuss the physical connection between this anomalous behavior and the gapless Dirac fermion fluctuations, corroborating that the dynamical feature of fermions in real-time case~\cite{Yin2019prl} persists in imaginary-time dynamics.
Moreover, our scaling theory allows precise determination of the critical exponents of the chiral Heisenberg Gross-Neveu transition, in agreement with previous equilibrium studies, but using only short-time data. The proposed framework not only provides a unified understanding of nonequilibrium ITCD in Dirac systems but also offers a practical and general route for investigating critical properties in strongly correlated systems, even enabling the efficient preemption of the sign problem in fermionic QMC simulations~\cite{yu2024sign}.

\begin{figure}[tbp]
\centering
  \includegraphics[width=0.9\linewidth]{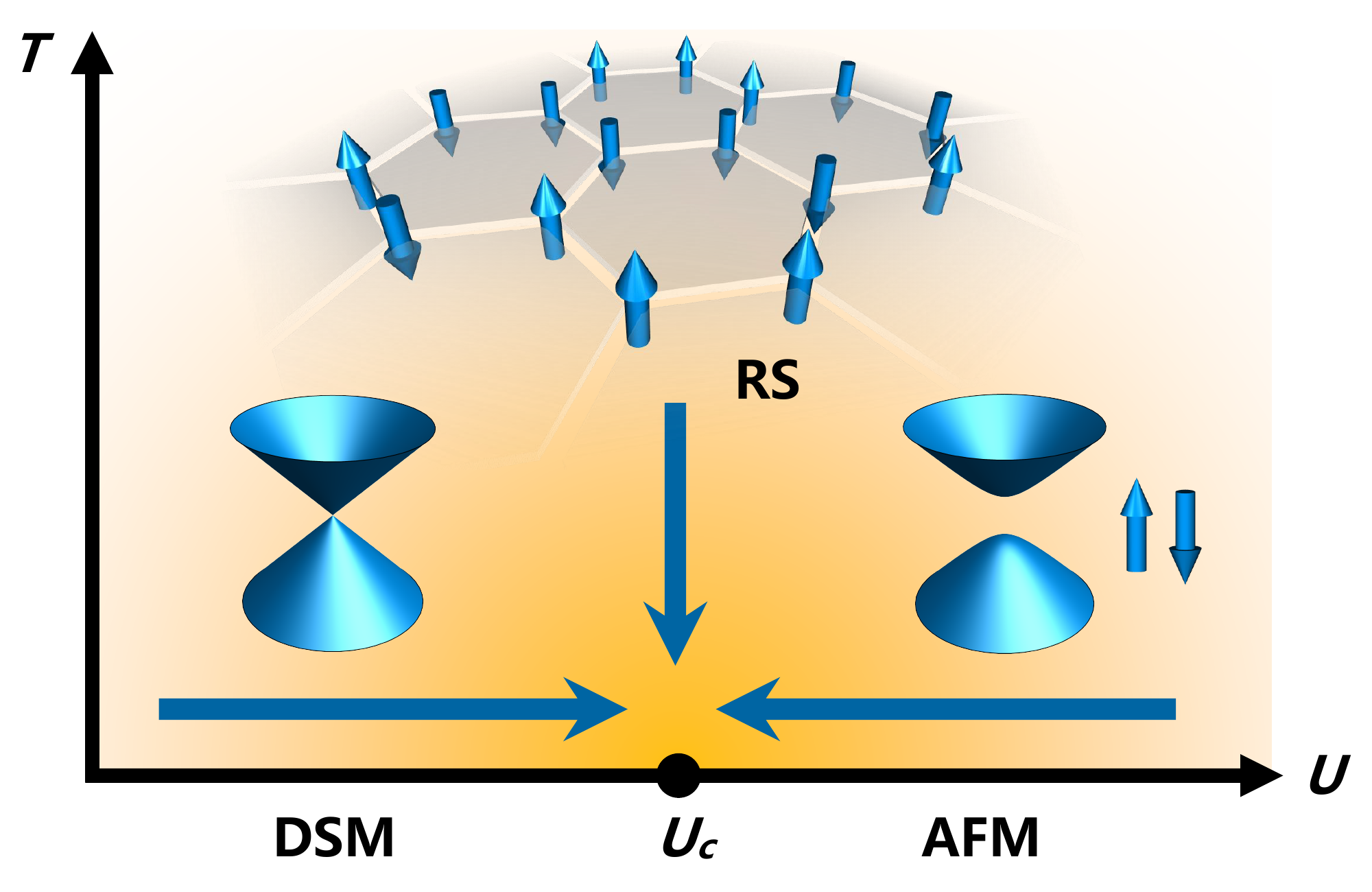}
  \vskip-3mm
  \caption{Sketch of the phase diagram and the quench protocol in imaginary-time with different initial states. The initial states are prepared as (i) the Dirac semimetal (DSM) phase, (ii) the saturated AFM state, and (iii) the random spin (RS) state. All states correspond to the fixed points of the initial states under the renormalization group transformation.
  }
  \label{figure1}
\end{figure}

\pdfbookmark[1]{Hamiltonian and dynamical protocol}{sec1}
\textcolor{black}{\it Hamiltonian and dynamical protocol}--- To explore the dynamic scaling in chiral Heisenberg universality class, we start with the Hubbard model defined on the honeycomb lattice, characterized by the Hamiltonian~\cite{Sorella1992,Herbut2006prl,Herbut2009prb,Herbut2013prx,Sorella2016prx}:
\begin{equation}
 H=-t\sum_{\langle ij\rangle,\sigma}c_{i\sigma}^\dagger c_{j\sigma}+U\sum_i \left({n_{i\uparrow}-\frac{1}{2}}\right) \left({n_{i\downarrow}-\frac{1}{2}}\right) \label{eq:Hamiltonian},
\end{equation}
in which $c_{i\sigma}^\dagger$ ($ c_{j\sigma}$) represents the creation (annihilation) operator of electrons with spin polarization $\sigma$, $n_{i\sigma}\equiv c_{i\sigma}^\dagger c_{i\sigma}$ is the electron number operator, $t$ is hopping amplitude between the nearest neighbor sites, and $U$ represents the strength of on-site repulsive interaction. As shown in Fig.~\ref{figure1}, when $U/t\ll 1$ the system is in the Dirac semimetal (DSM) phase characterized by the four-component Dirac excitation with flavor number $N_f=2$; whereas for large $U/t\gg 1$ the system hosts an antiferromagnetic (AFM) phase with a finite charge gap. A phase transition separating these two phases happens at a finite $U_c/t \approx 3.9$ and belongs to the chiral Heisenberg universality class~\cite{Sorella1992,Herbut2006prl,Herbut2009prb,Herbut2013prx,Sorella2016prx}. For simplicity, we set $t$ to unity in subsequent discussions.

For the imaginary-time relaxation dynamics, the wave function $|\psi(\tau)\rangle$ evolves according to the imaginary-time Schr\"{o}dinger equation $-\frac{\partial}{\partial \tau} |\psi(\tau)\rangle=H|\psi(\tau)\rangle$
imposed by the normalization condition. The formal solution of the Schr\"{o}dinger equation is given by $|\psi(\tau)\rangle=e^{-\tau H}|\psi(0)\rangle/Z(\tau)$, in which $Z(\tau)\equiv \langle \psi(\tau)|\psi(\tau)\rangle$ is the normalization factor and $|\psi(0)\rangle$ is the initial wavefunction. As illustrated in Fig.~\ref{figure1}, we will consider three kinds of initial states: (i) the saturated AFM state, (ii) the non-interacting DSM state, and (iii) the random spin (RS) state. In the following, we will employ the large-scale determinant QMC method to simulate the ITCD.

\pdfbookmark[1]{General scaling theory}{sec1}
\textcolor{black}{\it General scaling theory}--- Generally, for an observable $P$ its dynamic scaling in ITCD should satisfy~\cite{Yins2014prb,Zhengb1996prl}:
\begin{equation}
\label{eq:operator}
P\left(\tau,g,L,\{X\}\right)=\tau^{-\frac{\kappa}{z}}f_P\left(g \tau^\frac{1}{\nu z},L^{-1}\tau^{\frac{1}{z}},\{X \tau^{-\frac{c}{z}}\}\right),
\end{equation}
in which $g\equiv (U-U_c)$, $L$ is the lattice size, $\kappa$ is scaling dimension of $P$, $\nu$ is the correlation length exponent, $z$ is dynamic exponent, and $z=1$ for the Dirac QCP in Eq.~(\ref{eq:Hamiltonian}) (This value can be determined via ITCD as discussed in Supplemental Material (SM)~\cite{SM}). $\{X\}$ with its exponent $c$ represents the possible relevant variables associated with the initial state.

\nocite{Sorella2016prx,Sorella2019prb,Lang2019PRL,Vaezi2022PRL,Migdal1957JETP,AssaadReview,AssaadALF,Suzuki2005book,Assaad1997prb,Masatoshi1997qmc,Janssen2014prb,Janssen1989,Yins2014prb,Zhengb1996prl,Yins2014pre,Zhengb1998review,Shu2017prb}

Three remarks on Eq.~(\ref{eq:operator}) are listed as follows. 
(a) For $\tau\rightarrow \infty$, Eq.~(\ref{eq:operator}) recovers the usual finite-size scaling and $\{X\}$ becomes irrelevant. (b) All three kinds of initial states studied here, namely AFM, DSM, and RS states, correspond to three fixed points, respectively~\footnote{The random spin (RS) state remains invariant under coarse-graining or renormalization group transformations, and can thus be regarded as a stable fixed point.}. Thus, $\{X\}$ does not explicitly appear in Eq.~(\ref{eq:operator}). However, scaling functions $f_P$ vary for different initial states. (c) We mainly focus on the short-time scale $1 \ll \tau \ll L^z$ in the sense of a field theory. This is distinct from both the long-time scale $\tau \gg L^z$ required for reaching the ground state, and the microscopic ultraviolet time scale. In this window, the system enters a nonequilibrium universal scaling regime governed by the underlying QCP.

\begin{figure}[tbp]
\centering
  \includegraphics[width=\linewidth]{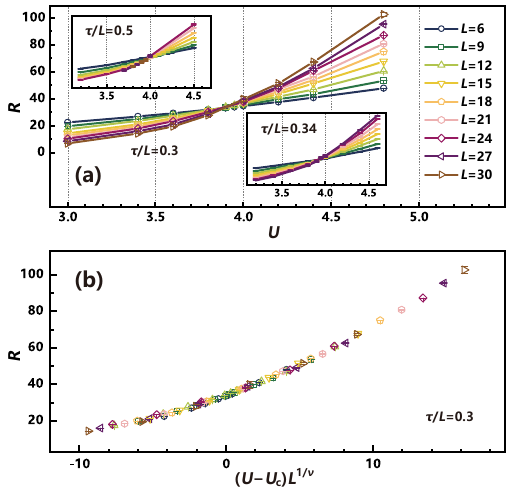}
  \vskip-3mm
  \caption{The results of correlation-length ratio $R$ against interaction $U$ for various sizes during the short-time stage, with a fixed value of $\tau L^{-z}$. (a) Estimation of the critical point via the intersection points of curves for $\tau L^{-z}=0.3$ (Main panel), $0.34$ and $0.5$ (Inset). (b) Estimation of $\nu$ by scaling collapse analysis of correlation-length ratio.
  }
  \label{figure2}
\end{figure}

\pdfbookmark[1]{Relaxation dynamics with AFM initial state}{sec1}
\textcolor{black}{\it Relaxation dynamics with AFM initial state}--- First, we study the ITCD starting with the AFM initial state. To illustrate the nonequilibrium scaling properties, we explore the dynamics of the correlation-length ratio defined as $R\equiv{S(\bm 0)}/{S(\Delta \bm q)}$~\cite{Assaad2015PRB}, where $\Delta\bm{q}$ is minimum lattice momentum and $S({\bm q})$ is the antiferromagnetic structure factor $S({\bm{q}})= \frac{1}{L^{2d}} \sum_{i,j} \mathrm{e}^{\mathrm{i} \bm{q}\cdot ({\bm{r}_i-\bm{r}_j})} \langle{S_i^z S_j^z}\rangle$ with $S_i^z$ being the staggered magnetization operator defined as $S_i^{z}\equiv\frac{1}{2}\bm{c}^\dagger_{i,A}\sigma^z\bm{c}_{i,A}-\frac{1}{2}\bm{c}^\dagger_{i,B}\sigma^z\bm{c}_{i,B}$ and ${\bm{c}^\dagger\equiv({c_\uparrow^\dagger,c_\downarrow^\dagger}})$.

As a dimensionless variable, $R$ in the ITCD obeys the following dynamic scaling form according to Eq.~(\ref{eq:operator}):
\begin{equation}
   R({g,\tau,L})=f_R(gL^{1/\nu},\tau L^{-z}), \label{eq:Rscale}
\end{equation}
which indicates that with a fixed $\tau L^{-z}$, $R$ does not depend on the system size when $g=0$, thereby providing a method to pinpoint the critical point.

As shown in Fig.~\ref{figure2} (a), we calculate $R$ as a function of $U$ with fixed $\tau L^{-z}=0.3$ for different sizes, and find that the curves almost cross at a point. Accordingly, we determine the critical point as $U_c=3.91(3)$ (See details in the SM~\cite{SM}). Upon fixing $U_c =3.91$ into Eq.~(\ref{eq:Rscale}), we adjust the value of $\nu$ for the rescaled horizontal variable $gL^{1/\nu}$ to make curves of different sizes collapse, yielding the value of $\nu$ as $\nu=1.17(7)$. Both values of $U_c$ and $\nu$ are consistent with those obtained from equilibrium method, albeit slight deviations arise possibly due to the scaling corrections~\cite{Sorella2016prx,Janssen2023prb}. Remarkably, significantly less effort is required as the results are obtained in the short-time stage and long enough imaginary-time evolution to achieve the ground state in the usual equilibrium method is not required here. Moreover, Eq.~(\ref{eq:Rscale}) also provides a self-consistent way to confirm the results. As shown in Fig.~\ref{figure2} and the SM~\cite{SM}, for different $\tau L^{-z}$, consistent $U_c$ and $\nu$ are obtained in a similar way, highlighting the validity of Eq.~(\ref{eq:Rscale}).
\begin{figure}[tbp]
\centering
  \includegraphics[width=\linewidth]{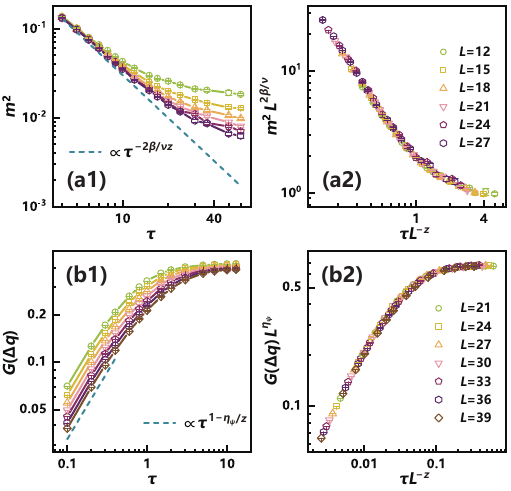}
  \vskip-3mm
  \caption{Relaxation dynamics at QCP with the AFM initial state. (a) Curves of $m^2$ versus $\tau$ for different sizes before (a1) and after (a2) rescaling. The dashed line in (a1) representing $m^2\propto \tau^{-2\beta/\nu z}$ with $\beta/\nu$ estimated from (a2) is plotted for comparison. (b) Curves of $G$ versus $\tau$ before (b1) and after (b2) rescaling. The dashed line in (b1) represents $G\propto \tau^{1-\eta_\psi/z}$ with $\eta_\psi$ estimated from (b2). 
  }
  \label{figure3}
\end{figure}

To delve deeper into the ITCD of model~(\ref{eq:Hamiltonian}), we study scaling behaviors of the square of order parameter $m^2=S(0)$. In the following, we focus on the case with $g=0$. The off-critical-point effects will be discussed in the SM~\cite{SM}. According to Eq.~(\ref{eq:operator}), for the saturated AFM initial state, $m^2$ should satisfy~\cite{Lizb1995prl,Yins2014prb}:
\begin{equation}
   m^2({\tau,L})=\tau^{-2\beta/\nu z}f_m(\tau L^{-z}), \label{eq:mscale}
\end{equation}
where $\beta/\nu$ is scaling dimension of $m$. Note that Eq.(\ref{eq:mscale}) can be transformed to $m^2({\tau,L})=L^{-2\beta/\nu}f_{m1}(\tau L^{-z})$~\footnote{\label{fn:scaling_transform} This transformation proceeds as follows. Starting from Eq.(\ref{eq:mscale}): $m^2(\tau, L) = \tau^{-2\beta/\nu z} f_m(\tau L^{-z})$, we define $f_m(\tau L^{-z}) = (\tau L^{-z})^{2\beta/\nu z} f_{m1}(\tau L^{-z})$. Substituting back gives $m^2(\tau, L) = \tau^{-2\beta/\nu z} (\tau L^{-z})^{2\beta/\nu z} f_{m1}(\tau L^{-z}) = L^{-2\beta/\nu} f_{m1}(\tau L^{-z})$.} so that the usual finite-size scaling is recovered as $\tau\rightarrow\infty$.

\begin{figure}[tb]
\centering
  \includegraphics[width=\linewidth]{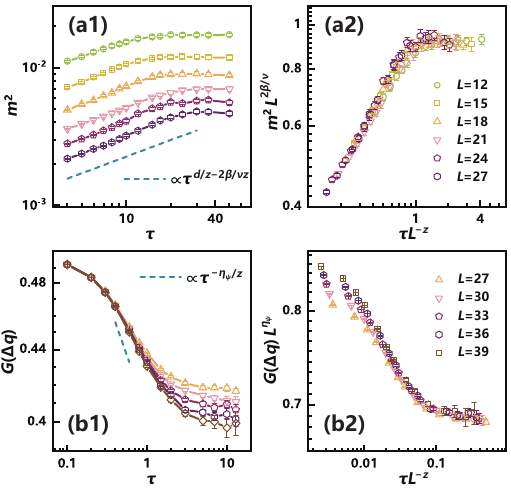}
  \vskip-3mm
  \caption{Relaxation dynamics at QCP with the DSM initial state. (a) Curves of $m^2$ versus $\tau$ at the critical point for different sizes before (a1) and after (a2) rescaling. The dashed line representing $m^2\propto \tau^{d/z-2\beta/\nu z}$ is plotted in (a1) for comparison. (b) Curves of $G$ versus $\tau$ before (b1) and after (b2) rescaling. The dashed line in (b1) represents $G\propto \tau^{-\eta_\psi/z}$. The critical exponents used here are estimated from Fig.~\ref{figure3}.
  }
  \label{figure4}
\end{figure}

Fig.~\ref{figure3} (a1) shows the evolution of $m^2$ for different $L$. At first, as shown in Fig.~\ref{figure3} (a2), data collapse analysis of the results yields the exponent $\beta/\nu = 0.80(3)$, which is close to the result obtained via equilibrium methods~\cite{Herbut2013prx,Sorella2016prx}.  The collapse of rescaled results for different $L$ into a single curve unequivocally demonstrates the dynamic scaling behavior with AFM as initial state, as depicted in Eq.~(\ref{eq:mscale}). Moreover, as shown in Fig.~\ref{figure3} (a1), one finds that in the short-time stage, $m^2\propto \tau^{-2 \beta/\nu z}$ and the scaling behavior is almost independent of the system size. The underlying reason is that the initial state is an uncorrelated state and the correlation length $\xi$ increases with time as $\xi\propto \tau^{1/z}$. In the short-time stage, $\xi< L$ and the finite-size effects are negligible; whereas in the long-time stage, $\xi> L$ and the system enters the finite-size scaling region in which $m^2\propto L^{-2 \beta/\nu}$. These results demonstrate that it is feasible to infer critical properties in the thermodynamic limit directly from the short-time dynamics.

Then we turn to the ITCD of the fermion correlation defined as $G(\Delta \bm q) \equiv \frac{1}{L^d} \sum_{ij} \mathrm{e}^{\mathrm{i} (K+\Delta \bm {q}) \cdot (\bm{r}_i - \bm{r}_j)} c_{i,A}^\dagger c_{j,B}$, with $K$ being the momentum at the Dirac point. According to Eq.~(\ref{eq:operator}), the scaling form of $G$ should be:
\begin{equation}
   G({\tau,L})=\tau^{-\eta_\psi/z}f_G(\tau L^{-z}), \label{eq:gscale}
\end{equation}
in which $\eta_\psi$ is the anomalous dimension of fermion operator. It was shown that $G$ is closely related to the quasi-particle weight $Z$~\cite{SM}, whose singularity was observed in the fermionic criticality~\cite{yang2023np}, also scales with $\eta_\psi$~\cite{SM}. Moreover, the fact that $G=0$ at the saturated AFM initial state dictates the absence of the constant term in the short-time expansion of $f_G(\tau L^{-z})$. Thus, Eq.~(\ref{eq:gscale}) can be transformed  into: 
\begin{equation}
   G({\tau,L})=\tau^{1-\eta_\psi/z}L^{-z}f_{G1}(\tau L^{-z}). \label{eq:gscale1}
\end{equation}

Fig.~\ref{figure3} (b1) shows the evolution of $G$ at $g=0$ for different $L$. Data collapse analysis in Fig.~\ref{figure3} (b2) gives $\eta_\psi = 0.15(4)$, which is close to the result obtained via equilibrium methods~~\cite{Sorella2016prx,Sorella2019prb}. Moreover, Fig.~\ref{figure3} (b1) shows that in the short-time stage, $G\propto \tau^{1-\eta_\psi/z}$. These results not only confirm Eq.~(\ref{eq:gscale1}), but also demonstrate that the critical exponent for fermion correlation can be determined from the ITCD.

\pdfbookmark[1]{Relaxation dynamics with DSM initial state}{sec1}
\textcolor{black}{\it Relaxation dynamics with DSM initial state}--- We proceed to explore the ITCD from the non-interacting DSM state. For this state, it is straightforward to show that $m^2\propto L^{-d}$. This size-dependent scaling affects the relaxation dynamics in the short-time stage, giving rise to the dynamic scaling of $m^2$:
\begin{equation}
   m^2({\tau,L})=L^{-d}\tau^{d/z-2\beta/\nu z}f_{m2}(\tau L^{-z}). \label{eq:mscale1}
\end{equation}

To demonstrate the dynamic scaling Eq.~(\ref{eq:mscale1}), we show in Fig.~\ref{figure4} (a) the evolution of $m^2$ at $g=0$. In the short-time stage, $m^2$ increases as $m^2\propto\tau^{d/z-2\beta/\nu z}$ for given $L$, qualitatively different from the dynamic behavior with AFM initial state. In addition, by rescaling $m^2$ and $\tau$ according to Eq.~(\ref{eq:operator}) with the exponents determined in previous section, one finds that the curves collapse onto each other. These results reveal the dynamic scaling behavior with DSM initial state described by Eq.~(\ref{eq:mscale1}).

\begin{figure}[tb]
\centering
  \includegraphics[width=\linewidth]{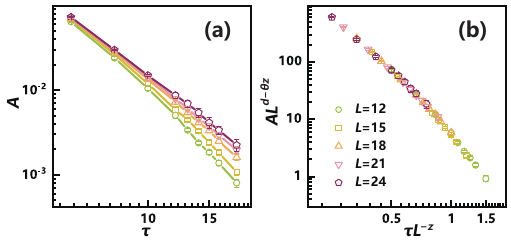}
  \vskip-3mm
  \caption{Critical initial slip manifested in the evolution of the auto-correlation function $A$ with the RS initial state. Curves of $A$ versus $\tau$ for different sizes at the critical point before (a) and after (b) rescaling.
  }
  \label{figure5}
\end{figure}

We also study the dynamics of fermion correlation $G$ as shown in  Fig.~\ref{figure4} (b). In the short-time stage, Fig.~\ref{figure4} (b1) shows that $G$ tends towards the curve of $G\propto \tau^{-\eta_\psi/z}$ as $L$ increases. Moreover, scaling collapse in Fig.~\ref{figure4} (b2) with exponents determined in previous section confirm Eq.~(\ref{eq:gscale}).

\pdfbookmark[1]{Critical initial slip with RS initial state}{sec1}
\textcolor{black}{\it Critical initial slip with RS initial state}--- Then we study the ITCD from the RS state, for which every site has one electron with its spin randomly distributed. With this uncorrelated initial state, the evolution of $m^2$ and $G$ satisfy Eq.~(\ref{eq:mscale1}) except for a different scaling function, as discussed in the SM~\cite{SM}.

Furthermore, when initiating from the RS state, we observe a universal critical initial slip behavior in the short-time stage~\cite{Janssen1989,Lizb1995prl,Yins2014prb}. Intriguingly, the scaling property of the initial slip behavior is determined by an independent dynamic exponent $\theta$, which does not exist in the equilibrium critical behavior. To characterize the critical initial slip, the auto-correlation function $A$ was introduced, defined as $A=\frac{1}{L^d}\sum_i\overline{\langle S_i(0)\rangle \langle S_i(\tau)\rangle}$~\cite{Zhengb1998review}, where the overline denotes the average on the initial state configurations. At the critical point, $A$ obeys the following scaling form~\cite{Zhengb1998review}:
\begin{equation}
   A(\tau)=\tau^{-d/z+\theta}f_A(\tau L^{-z}), \label{eq:auto}
\end{equation}
As shown in Fig.~\ref{figure5}, by rescaling $A$ and $\tau$ according to Eq.~(\ref{eq:auto}) and adjusting the trial value of the $\theta$ to make the rescaled curves collapse onto each other, we access the dynamic exponent $\theta=-0.84(4)$.

The negative exponent $\theta$ is notably interesting since it is in sharp contrast to classical criticality, where $\theta$ is typically positive. To illustrate the underlying physics, it is intuitive to consider the short-time dynamics of $m$ from an initial RS state with a small residual magnetization $m_0$. It was shown that $m$ evolves as $m \propto m_0 \tau^{\theta}$ in the short-time stage~\cite{Janssen1989}, governed by the competition between domain growth and critical fluctuations. 
In classical systems, critical fluctuations of the order parameter are not yet developed at early times, allowing magnetic domains to grow around the $m_0$ seeds and yielding $\theta>0$. This scenario also holds for the quantum Ising model~\cite{Yins2014prb,SM}, where bosonic order-parameter correlations remain weak in the short-time stage, and the $\theta>0$ critical initial slip originates from the mean-field tendency toward ordering~\cite{Yins2014prb}.
In contrast, in Dirac criticality, fermion correlations rapidly build up and equilibrate much earlier than their bosonic counterparts (see~\cite{SM}, Fig.~S4). This may arise from the linear dispersion and low density of states near the Dirac point, which lead to higher characteristic energies and shorter relaxation times than those of the bosonic modes. Consequently, the imaginary-short-time dynamics of the system is dominated by critical fluctuations of gapless fermions, while magnetic ordering is suppressed, resulting in a negative critical initial slip exponent $\theta<0$, in analogy to the real-time case~\cite{Yin2019prl}. 

\pdfbookmark[1]{Discussions and concluding remarks}{sec1}
\textcolor{black}{\it Discussions and concluding remarks}--- In summary, we perform sign-problem-free QMC simulation to investigate the ITCD in a Dirac QCP belonging to chiral Heisenberg universality class. For the first time, we extend the nonequilibrium scaling theory to Dirac systems and establish the corresponding scaling forms for different initial states, thereby revealing rich nonequilibrium dynamic scaling behaviors. Particularly, a negative critical initial slip exponent $\theta=-0.84(4)$ is observed in the ITCD from the RS initial state, remarkably different from the classical cases in which $\theta$ is usually positive, confirming the role of fermion fluctuations in relaxation dynamics~\cite{Yin2019prl}.

Our study also paves a new way to decipher the critical properties of quantum phase transitions in fermionic systems. Compared with the usual methods tackling critical properties in equilibrium ground state, the nonequilibrium method is highly efficient, since critical exponents are accessed by the short imaginary-time evolution. More crucially, our study offers a possible route to studying fermionic QCP in the presence of sign problem, which is the main obstacle hindering the understanding of QCP by numerical approach. Since the severity of sign problem exponentially increases with imaginary time in the process of evolution, the simulation on relatively large system sizes usually remains accessible in the stage of short imaginary time. Hence, it is promising to access the quantum critical behavior even when the model under consideration is sign problematic~\cite{yu2024sign}.

Moreover, inspired by the experimental realization of the ITCD in quantum computer platform~\cite{Zhang2023} and remarkable progresses made in fermionic quantum processors~\cite{Zoller2023pnas,Zoller2023fermion-qudit}, it is expected that our present work can be experimentally realized and contribute efficient approach to detect quantum criticality in these quantum devices in near future. In addition, it was shown that the driven critical dynamics and the shallow sudden quench critical dynamics are all described by the equilibrium critical exponents~\cite{zeng2024arxiv,Deng2011prb,Yin2019prl,Dziarmaga2010review,Polkovnikov2011rmp}. Accordingly, the critical exponents determined here are also applicable to these kinds of dynamics in Dirac criticality~\cite{zeng2024arxiv}.

\pdfbookmark[1]{Acknowledgments}{sec1}
\textcolor{black}{\it Acknowledgments}--- Y.-K. Y., Z. Z., and S. Y. are supported by the National Natural Science Foundation of China under Grants No. 12222515 and No. 12075324, the Research Center for Magnetoelectric Physics of Guangdong Province under Grant No. 2024B0303390001, and the Guangdong Provincial Key Laboratory of Magnetoelectric Physics and Devices under Grant No. 2022B1212010008. S. Y. is also supported by the Science and Technology Projects in Guangdong Province under Grant No. 2021QN02X561 and Guangzhou City under Grant No. 2025A04J5408. Z.-X. L. is supported by the start-up grant of IOP-CAS, the National Natural Science Foundation of China under Grants No. 12347107 and No. 12474146, the Beijing Natural Science Foundation under Grant No. JR25007, and the New Cornerstone Investigator Program. Y.-R. S. is supported by the National Natural Science Foundation of China under Grant No. 12104109. Y.-K. Y. and Z. Z. are also supported by the (national) college students innovation and entrepreneurship training program, Sun Yat-sen University.

\pdfbookmark[1]{Data availability}{sec1}
\textcolor{black}{\it Data availability}--- The data that support the findings of this article are openly available~\cite{yu2026zenodo}.

\bibliographystyle{apsrev4-1}
\bibliography{ref}

\onecolumngrid
\newpage
\widetext
\thispagestyle{empty}

\makeatletter
\let\addcontentsline\oldaddcontentsline
\makeatother

\setcounter{equation}{0}
\setcounter{figure}{0}
\setcounter{table}{0}
\renewcommand{\theequation}{S\arabic{equation}}
\renewcommand{\thefigure}{S\arabic{figure}}
\renewcommand{\thetable}{S\arabic{table}}

\renewcommand\floatpagefraction{0.9}
\renewcommand\textfraction{0.1}

\pdfbookmark[0]{Supplementary Materials}{SM}
\begin{center}
    \vspace{3em}
    {\Large\textbf{Supplementary Materials for}}\\
    \vspace{1em}
    {\large\textbf{Nonequilibrium dynamics in Dirac quantum criticality}}\\
    \vspace{0.5em}
\end{center}

\tableofcontents

\section{Determinant quantum Monte Carlo}

    We employ the large-scale determinant quantum Monte Carlo (DQMC) method~\cite{AssaadReview,AssaadALF} to investigate the imaginary-time relaxation dynamics of our model. Specifically, we prepare an initial state $\ket{\psi_0}$ and set the system parameters $U/t$ on the critical point to observe the scaling behavior of observables during the short-time stage. When the system evolves to imaginary time $\tau$, the expectation value of observables is given by
    \begin{equation}
        \braket{O(\tau)} = 
        \frac{
                \bra{\psi_0}\mathrm{e}^{-\frac{\tau}{2}H}~O~\mathrm{e}^{-\frac{\tau}{2}H}\ket{\psi_0}
            }{
                \bra{\psi_0}\mathrm{e}^{-\tau H}\ket{\psi_0}
            }.
    \end{equation}
    Herein, the imaginary-time propagator acts on the initial state, projecting it closer to the ground state. Hence, the DQMC framework under this context is also termed propagator quantum Monte Carlo (PQMC). In numerical calculations, we use Trotter decomposition to discretize imaginary-time propagator into ${M=\tau/\Delta\tau}$ (${M}$ is integer) time slices with~\cite{Suzuki2005book} 
    \begin{equation}
        \mathrm{e}^{-\tau H} = \prod^M_{m=1} \left[\mathrm{e}^{-\Delta\tau H_t} \mathrm{e}^{-\Delta\tau H_U} + \mathcal{O}\left(\Delta\tau^2\right)\right],
    \end{equation}
    where $H_t$ and $H_U$ are the hopping term and Hubbard interaction term respectively in the Hamiltonian. We choose small enough $\Delta\tau/t< 0.05$. To decouple two-body fermion-fermion coupling form of $\mathrm{e}^{\Delta\tau H_U}$, we use a discrete Hubbard-Stratonovich transformation~\cite{Assaad1997prb,Masatoshi1997qmc}
    \begin{equation}
        \mathrm{e}^{-\frac{\Delta\tau U}{2}\left(n_{i\uparrow}+n_{i\downarrow}-1\right)^2}
        =
        \sum_{l=\pm 1,\pm 2}
        \gamma (l)
        \mathrm{e}^{\mathrm{i} \sqrt{\frac{\Delta\tau U}{2}}\eta(l)\left(n_{i\uparrow}+n_{i\downarrow}-1\right)},
        \label{seq:HS}
    \end{equation}
    to obtain one-body fermion-auxiliary field coupling. Here, we introduce a four-component space-time local auxiliary fields $\gamma(\pm1)=1+\sqrt{6}/3$, $\gamma(\pm2)=1-\sqrt{6}/3$, $\eta(\pm1)=\pm\sqrt{2\left(3-\sqrt{6}\right)}$, $\eta(\pm2)=\pm\sqrt{2\left(3+\sqrt{6}\right)}$, and use DQMC for importance sampling over these space-time configurations. Next, we elaborate on how DQMC numerically calculates the sampling weight.

    For each imaginary time and each position of the Hubbard interaction, we employ an Hubbard-Stratonovich transformation as in Eq.~\eqref{seq:HS}. This means that we introduce an auxiliary field in $d+1$ dimensions. As a result, the imaginary-time propagator can be fully expressed using single-particle operators. This allows us to represent it in the following quadratic form of fermion operators:
    \begin{equation}
        \mathrm{e}^{-\tau H} 
        \equiv \sum_\mathrm{c} \mathrm{e}^{-\tau H_\mathrm{c}} 
        = \sum_\mathrm{c} A_{\mathrm{c}} \prod^M_{m=1} ~ \mathrm{e}^{ \vec{c}^\dagger T \vec{c}} ~ \mathrm{e}^{ \vec{c}^\dagger V_{\mathrm{c}(m)} \vec{c}},
        \label{seq:exp(-tH)}
    \end{equation}
    where $\sum_\mathrm{c}$ denotes the summation over all space-time configurations of the auxiliary field. Considering that each local component of the auxiliary field has 4 possible values, the summation comprises up to $4^{MN}$ terms, where $N$ represents the number of spatial degrees of freedom. $H_c$ denotes the decoupled configuration Hamiltonian, while $T$ and $V_{\mathrm{c}(m)}$ are the resulting quadratic coefficient matrices from the rearrangement, and $A_{\mathrm{c}}$ is the coefficient. Both $V_{\mathrm{c}(m)}$ and $A_{\mathrm{c}}$ depend on the auxiliary field configuration.
    The complete form of the evolution operator has been presented above. Next, we consider expressing the initial state. The AFM, DSM, RS initial states we use are all direct product states, and numerically they can be written as the following Slater determinant:
    \begin{equation}
        \ket{\psi_0} 
        = \bigotimes^{N_\mathrm{e}}_{n_\mathrm{e}=1} \left[ \left(\sum_x c_x^\dagger P_{x,n_\mathrm{e}}\right) \ket{0} \right]
        = \bigotimes^{N_\mathrm{e}}_{n_\mathrm{e}=1} \left[ \left( \vec{c}^\dagger P\right)_{n_\mathrm{e}} \ket{0} \right],
        \label{seq:|psi_0>}
    \end{equation}
    where $N_\mathrm{e}$ denotes the number of electrons. This implies that the initial state is a direct product of $N_\mathrm{e}$ fermion single-particle wave functions. The index $x$ denotes the degree of freedom of the electron, including spatial degrees of freedom, spin degrees of freedom, etc. The matrix element $P_{x,n_\mathrm{e}}$ represents the probability amplitude of the $n_\mathrm{e}$th electron on the $x$th degree of freedom.
    Note that the imaginary-time propagator $\mathrm{e}^{-\tau H_\mathrm{c}}$ is essentially the Boltzmann weight factor of the auxiliary field configuration in statistical mechanics. According to Eqs. \eqref{seq:exp(-tH)} and \eqref{seq:|psi_0>}, the partition function of the auxiliary field configuration can be expressed as:
    \begin{equation}
        Z 
        = \sum_\mathrm{c} \bra{\psi_0} \mathrm{e}^{-\tau H_\mathrm{c}} \ket{\psi_0} 
        = \sum_\mathrm{c} A_{\mathrm{c}} ~\mathrm{det} \left[P^\dagger B_{\mathrm{c}}(\tau,0)P\right].
        \label{seq:det}
    \end{equation}
    Here, we use $B_{\mathrm{c}}$ to represent the exponential of the quadratic coefficient matrix:
    \begin{equation}
        B_{\mathrm{c}}(\tau_2,\tau_1) \equiv \prod^{\tau_2/\Delta\tau}_{m=\tau_1/\Delta\tau} ~ \mathrm{e}^{T} ~ \mathrm{e}^{V_{\mathrm{c}(m)}}.
    \end{equation}
    The expression on the right side of Eq.~\eqref{seq:det} has integrated out the fermion operators, replacing the Grassmann numbers and fermion statistics, with a determinant representation that is computationally tractable. All matrix operations can be performed directly on a computer.
    
    Ultimately, our Monte Carlo sampling is conducted over space-time configurations. Numerically, the weight of a space-time configuration is $A_{\mathrm{c}} ~\mathrm{det} \left[P^\dagger B_{\mathrm{c}}(\tau,0)P\right]$.
    Following the classical Markov importance sampling method, we continuously make tentative flips to the local components of this $d+1$ dimensional auxiliary field. We then employ the Metropolis algorithm to calculate the probability of accepting these changes based on the ratio of configuration weights before and after the flip. Specifically, we need to compute the following weight ratio:
    \begin{equation}
        R_{\mathrm{c'c}} \equiv \frac{
            A_{\mathrm{c'}} ~\mathrm{det} \left[P^\dagger B_{\mathrm{c'}}(\tau,0)P\right]
        }{
            A_{\mathrm{c}} ~\mathrm{det} \left[P^\dagger B_{\mathrm{c}}(\tau,0)P\right]
        },
    \end{equation}
    where c$'$ represents the flipped configuration and c represents the original configuration. In fact, we do not need to compute the weights of the two configurations separately. This is because the flipping we perform is localized in space-time, so
    \begin{equation}
        B_{\mathrm{c'}}(\tau,0) = 
        B_{\mathrm{c}}(\tau,\zeta) \left(\mathbb{1}+\Delta_{\mathrm{c'c}}\right) 
        B_{\mathrm{c}}(\zeta,0).
    \end{equation}
    Here, $\Delta_{\mathrm{c'c}}$ is a highly sparse matrix, where only the matrix elements corresponding to the degrees of freedom involved in the local auxiliary field flipping are non-zero. Thus, the ratio of the two determinants can be expressed as:
    \begin{equation}
        \frac{\mathrm{det}\left[P^\dagger B_{\mathrm{c'}}(\tau,0)P\right]}{\mathrm{det}\left[P^\dagger B_{\mathrm{c}}(\tau,0)P\right]} = \mathrm{det} \left\{ 
            \mathbb{1}+\Delta_{\mathrm{c'c}} 
            B_{\mathrm{c}}(\zeta,0)P 
            \left[P^\dagger B_{\mathrm{c}}(\tau,0)P\right]^{-1}
            P^\dagger B_{\mathrm{c}}(\tau,\zeta)
        \right\}.
    \end{equation}
    Due to the sparsity of $\Delta_{\mathrm{c'c}}$, the determinant on the right side of the above equation only requires consideration of a few degrees of freedom involved in the flipping during computations.

    In DQMC, to compute the physical observables, we only need to statistically analyze the configurational observable $\braket{O(\tau)}_\mathrm{c}$.
    \begin{equation}
        \braket{O(\tau)} = \sum_\mathrm{c} \mathrm{Pr_c} \braket{O(\tau)}_\mathrm{c} + \mathcal{O}\left(\Delta\tau^2\right),
    \end{equation}
    where $\mathrm{Pr_c}$ represents the configuration probability,
    \begin{equation}
        \mathrm{Pr_c} = \frac{1}{Z} A_{\mathrm{c}} ~\mathrm{det} \left[P^\dagger B_{\mathrm{c}}(\tau,0)P\right],
    \end{equation}
    \begin{equation}
        \braket{O(\tau)}_\mathrm{c} = 
        \frac{
                \bra{\psi_0}\mathrm{e}^{-\frac{\tau}{2}H_{\mathrm{c}}}~O~\mathrm{e}^{-\frac{\tau}{2}H_{\mathrm{c}}}\ket{\psi_0}
            }{
                \bra{\psi_0}\mathrm{e}^{-\tau H_{\mathrm{c}}}\ket{\psi_0}
            }.
    \end{equation}
    Since we employ importance sampling, the sampling frequency is proportional to the configuration probability. Ultimately, when calculating the observable, we simply take the average over the sampled configurational observables. If the observable is a single-particle operator, meaning it can be expressed as a quadratic form of fermion operators, then one can integrate out the fermion degrees of freedom in a manner similar to Eq. \eqref{seq:det} and numerically compute using determinants. For observables of four-fermion operators or higher, we compute using the fermion equal-time Green's function based on Wick's theorem. After numerically integrating out the fermion degrees of freedom, the fermion equal-time Green's function can be expressed using the following matrix element:
    \begin{equation}
        \braket{c_{x_1}^\dagger c_{x_2}}_\mathrm{c} = 
        \left\{  
            B_{\mathrm{c}}\left(\frac{\tau}{2},0\right)P 
            \left[P^\dagger B_{\mathrm{c}}(\tau,0)P\right]^{-1}
            P^\dagger B_{\mathrm{c}}\left(\tau,\frac{\tau}{2}\right)
        \right\}_{x_1,x_2}.
    \end{equation}

\section{Determination of the critical point}

\begin{figure}[htbp]
        \centering
        \includegraphics{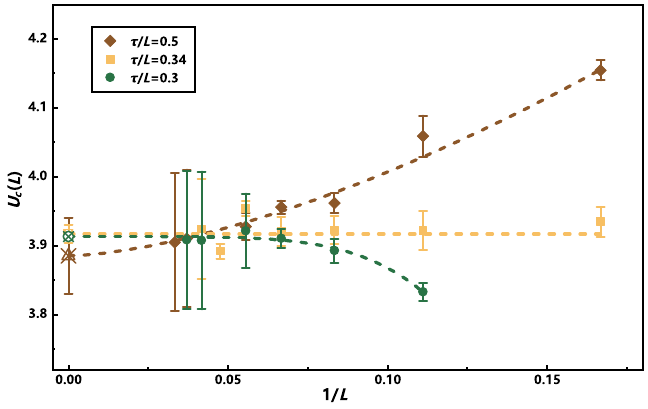}
        \caption{Determine the critical point by extrapolation. When $\tau L^{-z}$ is fixed at 0.3, 0.34, and 0.5 respectively, the intersection points $U_c(L)$ of the curves of $R$ versus $U$ for sizes $L$ and $L+3$ are shown in the figure. The critical point is extrapolated when the system tends to thermodynamic limit $L\to\infty$. Note that the extrapolation results are also marked at $1/L=0$.}
        \label{sfig:AFM_Uc_L}
    \end{figure}

    Here we offer supplementary details and numerical insights on pinpointing the critical point. In Fig. 2 (a) of the main text, we show that curves of $R$ versus $U$ for different sizes intersect at the critical point when $\tau L^{-z}$ is taken as $0.3$, $0.34$, and $0.5$. However, due to finite size effects, there may be slight deviations between the intersection points of small-sized curves and the real critical point. We denote the intersection points of size $L$ and $L+3$ as $U_c(L)$, which are shown in Fig.~\ref{sfig:AFM_Uc_L}.

    We extrapolate the critical point in thermodynamic limit $L\to\infty$ using the form $U_c(L)=U_c+aL^{-w}$. The intercept shown in Fig.~\ref{sfig:AFM_Uc_L} represents the extrapolated critical point. For three different cases of $\tau L^{-z}=0.3$, $0.34$, and $0.5$, our results are $U_c=3.91(3)$, $3.92(1)$, and $3.88(4)$ respectively. They all extrapolate to the same limit within the error range, thus demonstrating that our method based on nonequilibrium information to determine critical point in fermion criticality is reliable.

    It is worth mentioning that as shown in Fig.~\ref{sfig:AFM_Uc_L}, the $U_c(L)$ approaches the critical point with different trends as $L$ increases for different values of ${\tau/L}$. Specifically, when ${\tau/L=0.3}$, the $U_c(L)$ under small size is smaller than the real critical point at thermodynamic limit; while when ${\tau/L=0.5}$, the $U_c(L)$ under small size is larger than the real critical point. More interestingly, we found that at ${\tau/L=0.34}$, the $U_c(L)$ almost does not depend on size and can exhibit a real critical behavior under small sizes alone. These results demonstrate that our method is quite reliable in determining the critical point.

\section{More results about the critical exponents}
\subsection{The correlation length exponent $\nu$}
\begin{figure}[htbp]
        \centering
        \includegraphics{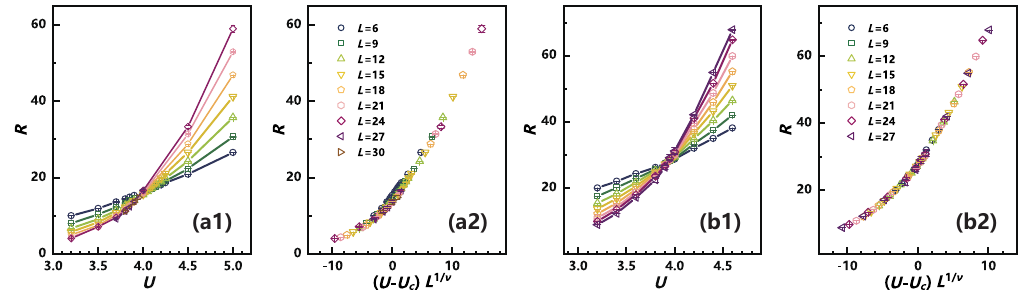}
        \caption{The variation of the correlation-length ratio with respect to $U/t$ at fixed $\tau L^{-z}$. (a1) display the results obtained by setting $\tau L^{-z}=0.5$. (a2) shows rescaling is applied to the horizontal axis of (a1). (b1) display the results obtained by setting $\tau L^{-z}=0.34$. (b2) shows rescaling is applied to the horizontal axis of (b1).}
        \label{sfig:AFM_nu_0.5and0.34}
    \end{figure}

    In the main text, we use the data with fixed $\tau L^{-z}=0.3$ to determine the critical exponent $\nu$. Here, we supplement the results of $\tau L^{-z}=0.5$ and $0.34$. Fig.~\ref{sfig:AFM_nu_0.5and0.34} shows the curves of different sizes' correlation-length ratios changing with $U/t$. We adjust the rescaling parameters $U_c$ and $\nu$ to make the curves of different sizes coincide. To avoid finite size effects as much as possible, we fit $U_c$ and $\nu$ using curves above $L=12$. For $\tau L^{-z}=0.5$ and $0.34$, we obtain results of $U_c=3.91(2)$ and $3.92(4)$, respectively, as well as $\nu=1.22(6)$ and $1.22(17)$. They are consistent with our results presented in the main text, and consistent with the previous results of equilibrium systems within error range~\cite{Janssen2014prb}.

    Moreover, by using these results for rescaling, in Fig.~\ref{sfig:AFM_nu_0.5and0.34} (a2), we find that for $\tau L^{-z}=0.5$, curves for small $L$ do not completely coincide with large-size curves; while in Fig.~\ref{sfig:AFM_nu_0.5and0.34} (b2), it is shown that all curves can coincide very well for $\tau L^{-z}=0.34$. These results are consistent with the drift of $U_c(L)$ found in the previous section.
    
\subsection{The dynamic exponent $z$}
    In the main text, we directly set the dynamic exponent $z$ as $z=1$, which is based on the Lorentz symmetry of effective theory of the Dirac criticality. Here we show that the dynamic exponent $z$ can be determined independently.
    
    \begin{figure}[htbp]
        \centering
        \includegraphics{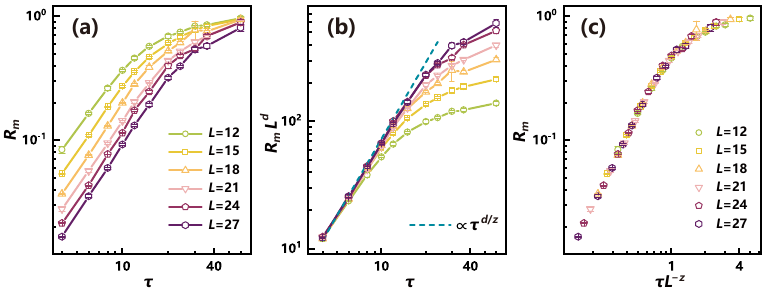}
        \caption{Evolution of $R_m$ for different sizes $L$. (a) Curves of $R_m$ versus $\tau$ at the critical point for different sizes (b) For small $\tau$, $R_m$ obeys $R_m L^d \sim \tau^{d/z}$ for different sizes. The dashed line with $z=1$ is plotted for comparison. (c) By rescaling the imaginary time $\tau$ as $\tau L^{-z}$, data collapse fitting yields $z=1.01(9)$.}
        \label{sfig:z}
    \end{figure}

    To do this, we define the ratio of $m^2$ for the DSM and AFM initial states as $R_m\equiv m^2_\mathrm{DSM}/m^2_\mathrm{AFM}$. At the critical point, according to Eqs. (4) and (7) in the main text, one finds that $R_m$ satisfies the scaling form
    \begin{equation}
        R_m(\tau,L)
        = L^{-d}\tau^{d/z}f_{R_m}(\tau L^{-z}).
        \label{seq:m/m}
    \end{equation}
    In this scaling function, $d=2$ and $z$ is the only independent critical exponent.
    
    Based on Eq.~(\ref{seq:m/m}), there are two approaches to determine $z$: 
    
    (a) In the short-time stage, $f_{Rm}$ only weakly depends on $L$. So, $R_m$ in the short-time stage should satisfy
    \begin{equation}
        R_m(\tau,L)\propto L^{-d}\tau^{d/z}.
        \label{seq:m/m1}
    \end{equation}
    Figure~\ref{sfig:z} (a) shows the evolution of $R_m$ for different size $L$. By rescaling $R_m$ as $R_m L^d$, Fig.~\ref{sfig:z} (b) confirms that $f_{Gm}$ only weakly depends on $L$ in the short-time stage. Then, we directly fit the curve for largest size in the short-time range according to Eq.~(\ref{seq:m/m1}) and find that $z=1.07(1)$, which is close to the exact result with $z=1$.    
    
    (b) A more accurate method is to adjust the value of $z$ for the rescaled curves of $R_m=L^{-d}\tau^{d/z}$ versus $\tau L^{-z}$ according to Eq.~(\ref{seq:m/m}) to make curves of different sizes collapse with each other, yielding the value of $z$ as $z=1.01(9)$, as shown in Fig.~\ref{sfig:z} (c).
    
   Although for the model studied in this paper, the dynamic exponent $z=1$ can be obtained from the Lorentz symmetry of the effective Hamiltonian, determining $z$ independently is also crucial. First, we can examine the effective field theory for the microscopic model. Second, in estimation of the critical point, we fix $\tau L^{-z}$ to be a constant with $z=1$. Determining $z$ at the critical point gives a consistent examination of the procedure.

   \section{Dynamic scaling of $m^2$ and $G$ from random-spin initial state}

   In the main text, for the random-spin (RS) initial state, we only focus on the critical initial slip behaviors. Here we investigate the imaginary-time critical dynamics of the square of the order parameter $m^2$ and the fermion correlation function $G$.

\begin{figure}[htbp]
        \centering
        \includegraphics{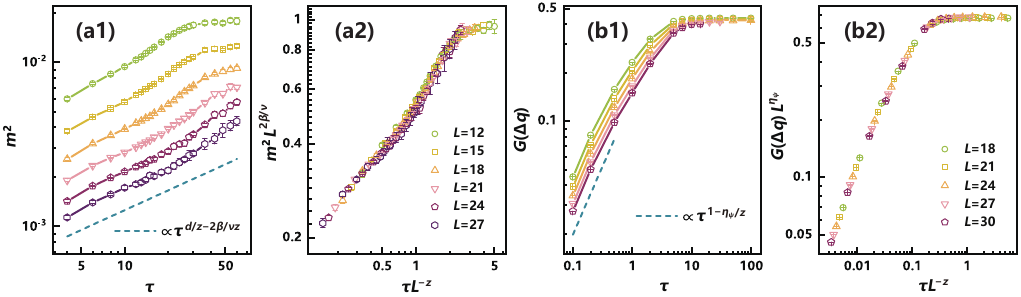}
        \caption{Relaxation dynamics of order paremeter $m^2$ and fermion correlation $G(\Delta \bm{q})$ with the RS initial state. Curves of $m^2$ versus $\tau$ at the critical point for different sizes before (a1) and after (a2) rescaling. The dashed line in (a1) representing $m^2\propto \tau^{-2\beta/\nu z}$ with $\beta/\nu=0.80$ estimated from the main text is plotted for comparison. Curves of $G(\Delta \bm{q})$ versus $\tau$ at the critical point for different sizes before (b1) and after (b2) rescaling. The dashed line in (b1) representing $G(\Delta \bm{q})\propto \tau^{1-2\eta_\psi/z}$ with $\eta_\psi=0.15(4)$ estimated from the main text is plotted for comparison.}
        \label{sfig:RS_m2_G}
    \end{figure}

  At first, we show that $m^2$ obeys
  \begin{equation}
   m^2({\tau,L})=L^{-d}\tau^{d/z-2\beta/\nu z}f_{m3}(\tau L^{-z}), \label{eq:mscalers}
  \end{equation}
  which is similar to Eq.~(7). This is because both RS and DSM states are disordered for the magnetic order. Figure~\ref{sfig:RS_m2_G} (a1) shows the evolution of $m^2$ for different $L$. From Fig.~\ref{sfig:RS_m2_G} (a1), one finds that in the short-time stage, $m^2\propto L^{-d}\tau^{d/z-2\beta/\nu z}$. By rescaling $m^2$ and $\tau$ as $m^2 L^{2\beta/\nu z}$ and $\tau L^{-z}$, respectively, with the critical exponents determined in the main text, Fig.~\ref{sfig:RS_m2_G} (a2) shows that the rescaled curves collapse onto a single curve, confirming Eq.~(\ref{eq:mscalers}) and the values of the critical exponents.

    Then, we turn to explore the scaling behavior of $G(\Delta \bm{q})$. We find that $G$ satisfies
    \begin{equation}
        G({\tau,L})=L^{-z}\tau^{1-\eta_\psi/ z}f_{G2}(\tau L^{-z}), \label{eq:grs}
    \end{equation}
    which is similar to Eq.~(7). This is because both RS and Mott insulator states are gapped phase for the fermion particles. Figure~\ref{sfig:RS_m2_G} (b1) shows the evolution of $G$ for different $L$. In the short-time stage, the curve of $G$ versus $\tau$ is consistent with $G\propto \tau^{1-\eta_\psi/z}$. Moreover, by rescaling $G$ and $\tau$ as $G L^{\eta_\psi}$ and $\tau L^{-z}$, respectively, with $\eta_\psi=0.15(4)$ determined in the main text, we find in Fig.~\ref{sfig:RS_m2_G} (b2) that the rescaled curves collapse onto each other, confirming not only Eq.~(\ref{eq:grs}) but also the value of $\eta_\psi$.
    
    Combining the scaling forms of the square of the order parameter $m^2$ and $G$, we find interesting results that scaling forms of $m^2$ for the DSM and RS initial states are similar; whereas scaling forms of $G$ for AFM and RS initial states are similar. These results provide further crosscheck for the critical exponents determined via nonequilibrium methods.

\section{Critical initial slip}
\subsection{Critical initial slip in the order parameter}

In the main text, we unveil the critical initial slip via the scaling behavior of the auto-correlation function with the RS initial state. Actually, the critical initital slip exponent also manifests itself in the evolution of the order parameter $m$~\cite{Janssen1989,Yins2014prb}. 

To exhibit the critical initial slip in $m$, the initial state should be changed to the uncorrelated initial state with a very small magnetization $m_0$. Under this condition, $m$ evolves as~\cite{Janssen1989,Yins2014prb}
\begin{equation}
        m = \tau^{-\frac{\beta}{\nu z}} f_{m4} (m_0 \tau^{\theta+\beta/\nu z}).
        \label{seq:m0}
\end{equation}
Here we at first focus on the thermodynamic limit. Expanding $m_0$ in $f_{m4}$ gives the leading term of $m$ as
\begin{equation}
        m\propto m_0\tau^{\theta}.
        \label{seq:m01}
\end{equation}

When $m_0$ becomes larger, the rescaled scaling variable $m_0 \tau^{\theta+\beta/\nu z}$ becomes invalid. Instead, a universal characteristic function should be introduced. For this case, the evolution in the initial stage no longer demonstrates power behavior~\cite{Zhengb1996prl,Yins2014pre}. These scaling behaviors have been found in both classical relaxation dynamics and quantum imaginary-time relaxation dynamics.

For classical systems, in general, $\theta$ is positive. Thus, in the initial stage, the order parameter increases, rather than directly decays to its equilibrium value $m=0$. The reason for this critical initial slip behavior is the competition between the domain formation and the critical fluctuations. The former describes the tendency that the local spin always wants to make its neighbors orient along the same (for ferromagnetic case)/inverse (for antiferromagnetic) direction, thus forming the magnetic domain; while the latter describes the destruction of the order by strong fluctuations near the critical point. In equilibrium, the latter dominates and $m=0$ at the critical point. However, since the initial state is an uncorrelated state, from which critical fluctuations should increase from infancy. Thus, in the early stage, the domain formation dominates and $m$ increases. As time goes on, the critical fluctuations becomes strong enough to destroy the order, and $m$ begin to decay as $m\propto \tau^{-\beta/\nu z}$.

However, in the present case, from Fig.~\ref{sfig:RS_m2_G}, we find that the fermion correlation function always tends to its equilibrium value much faster than the order parameter. Thus, gapless Dirac fermions can contribute significant critical fluctuations to prevent the domain formation, making $\theta$ negative.

Although $m$ also contains the information of $\theta$, here it is very difficult to obtain $\theta$ from the scaling of $m$ in Monte Carlo simulation. The reason is that the scaling relation $m\propto m_0\tau^{\theta}$ requires that $m_0$ should be very small. For finite-size systems employed in our manuscript, the smallest $m_0$ is $m_0\sim L^{-2}$, which is too large to satisfy the condition. Moreover, for finite-size system, $L$ should also be considered as a scaling variable. So, in our paper, we use the auto-correlation function from a random initial state to get the value of $\theta$. This is also a usual method to obtain $\theta$ in classical systems~\cite{Zhengb1998review}.

\subsection{$\theta$ for quantum Ising models}

\begin{figure}[htbp]
        \centering
        \includegraphics{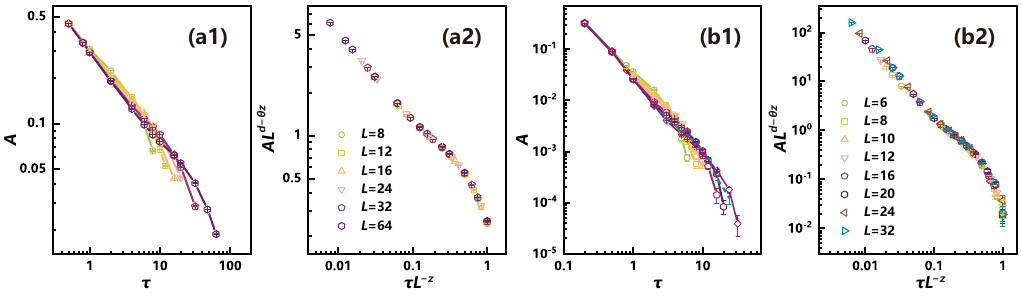}
        \caption{The relaxation behavior of the auto-correlation function $A$ in the quantum Ising model. (a1) and (b1) respectively show the results in 1D and 2D. (a2) and (b2) are their rescaled results, where the relaxation curves of different sizes overlap.}
        \label{sfig:TFIM_theta}
    \end{figure}

    In the main text, we determine the critical initial slip exponent $\theta$ of the Dirac fermions through the critical relaxation behavior of the auto-correlation function $A$. Here, we study the critical dynamics of the auto-correlation function $A$ in the $1$D and $2$D transverse-field Ising models. In previous works~\cite{Yins2014prb,Shu2017prb}, the critical initial slip exponent for the $1$D and $2$D Ising models is obtained from other methods. Here, we show that the critical initial slip exponent for these models can also be obtained from the scaling of $A$. Fig.~\ref{sfig:TFIM_theta} (a1) shows the relaxation process of $A$ for various sizes of the $1$D transverse-field Ising model at the critical point $h/J=1$. In Fig. \ref{sfig:TFIM_theta} (a2), we rescale the relaxation process for $1$D. Here, we take $\theta=0.3734$ \cite{Shu2017prb} and $z=1$. After rescaling, the relaxation curves of different sizes overlap, satisfying the scaling relation for $A$ as mentioned in our main text. For the $2$D transverse-field Ising model at the critical point $h/J=3.04451$, we performed similar numerical simulations, as shown in Fig. \ref{sfig:TFIM_theta} (b1). The relaxation curves of various sizes of $A$ overlap when rescaled with $\theta=0.209$ \cite{Shu2017prb}, as shown in Fig. \ref{sfig:TFIM_theta} (b2). Note that for both $1$D and $2$D quantum Ising models, the critical initial slip exponent is positive.

\section{Off-critical-point effects}
    
    In the main text, after determining the critical point, we focus on the case with $g=0$. Here, we discuss the off-critical-point effects.

    \subsection{Scaling forms with $g\ne0$}
    When $g\ne0$, the scaling functions should include $gL^{1/\nu}$ as an indispensable variable. Therefore, for the AFM initial state, the evolution of $m^2$ should satisfy  
    \begin{equation}
        m^2({g,L,\tau})=\tau^{-2 \beta/\nu z}f_{g1}(\tau L^{-z},gL^{1/\nu}),
        \label{aaa}
    \end{equation}
    which can be transformed to be $m^2({g,L,\tau})=L^{-2 \beta/\nu}f_{g2}(\tau L^{-z},gL^{1/\nu})$. Similarly, with the AFM initial state, the fermion correlation function $G$ should satisfy
    \begin{equation}
        G({g,L,\tau})=L^{-z}\tau^{1-\eta_\psi/ z}f_{g3}(\tau L^{-z},gL^{1/\nu}),
        \label{bbb}
    \end{equation}
    which is equivalent to $G({g,L,\tau})=L^{-\eta_\psi}f_{g4}(\tau L^{-z},gL^{1/\nu})$ according to the scaling transformation.

    In addition, for the DSM initial state, the evolution of $m^2$ should satisfy  
    \begin{equation}
        m^2({g,L,\tau})=L^{-d}\tau^{d/z-2 \beta/\nu z}f_{g5}(\tau L^{-z},gL^{1/\nu}),
        \label{ccc}
    \end{equation}
    which can be transformed to be $m^2({g,L,\tau})=L^{-2 \beta/\nu}f_{g6}(\tau L^{-z},gL^{1/\nu})$. Similarly, with the DSM initial state, the fermion correlation function $G$ should satisfy
    \begin{equation}
        G({g,L,\tau})=L^{-z}\tau^{1-\eta_\psi/ z}f_{g7}(\tau L^{-z},gL^{1/\nu}),
        \label{ddd}
    \end{equation}
    which is equivalent to $G({g,L,\tau})=L^{-\eta_\psi}f_{g8}(\tau L^{-z},gL^{1/\nu})$. 

\subsection{Numerical Results}
    
\begin{figure}[htbp]
        \centering
        \includegraphics{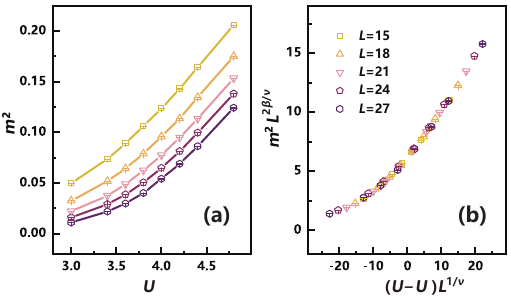}
        \caption{Estimation of $\nu$ and $\beta$ by scaling collapse analysis of $m^2$. Curves of $m^2$ versus $U$ with fixed $\tau L^{-z}=0.3$ before (a) and after (b) rescaling.}
        \label{sfig:AFM_m2U}
    \end{figure}

    At first, we show the numerical results for the AFM initial state. For a fixed $\tau L^{-z}$, the scaling form of $m^2$ reduces to $m^2({g,L})=L^{-2 \beta/\nu}f_{m2}(gL^{1/\nu})$. Fig.~\ref{sfig:AFM_m2U} depicts the dependence of $m^2$ on $U$ for different system sizes at $\tau L^{-z}=0.3$. By tuning the exponents to make the rescaled curves of $m^2$ versus $g$ collapse, we determine the exponents as $\nu=1.025(9)$ and $\beta/\nu=0.735(2)$, as shown in Fig.~\ref{sfig:AFM_m2U} (b). These values are consistent with the previous results of equilibrium systems \cite{Sorella2016prx} and also consistent with the non-equilibrium results determined at QCP in our main text.
    
    \begin{figure}[htbp]
        \centering
        \includegraphics{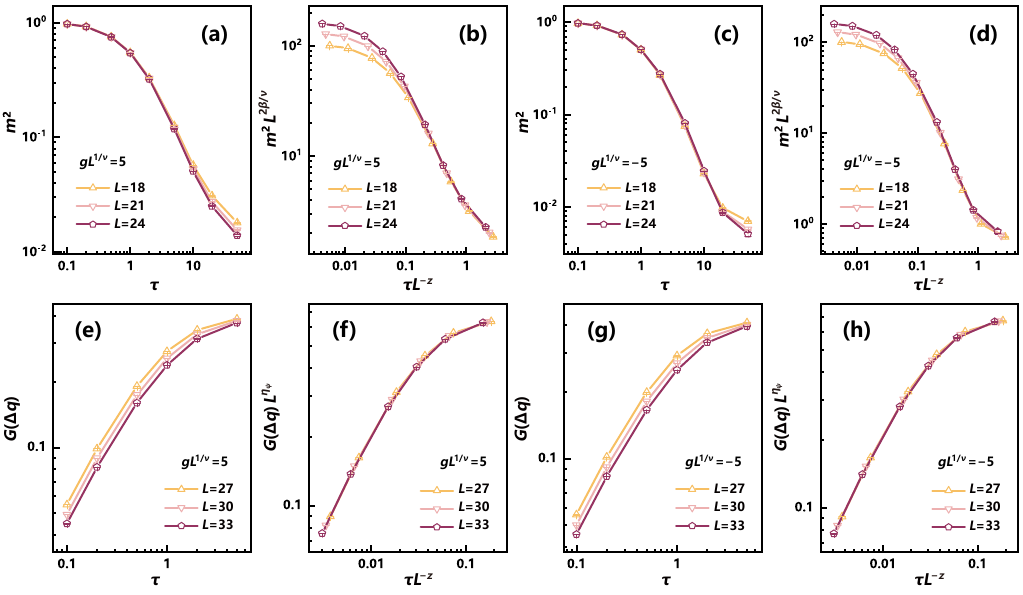}
        \caption{Relaxation dynamics of $m^2$ and $G$ from AFM initial state for $g\ne 0$.}
        \label{sfig:away_AFM}
    \end{figure}
    \begin{figure}[htbp]
        \centering
        \includegraphics{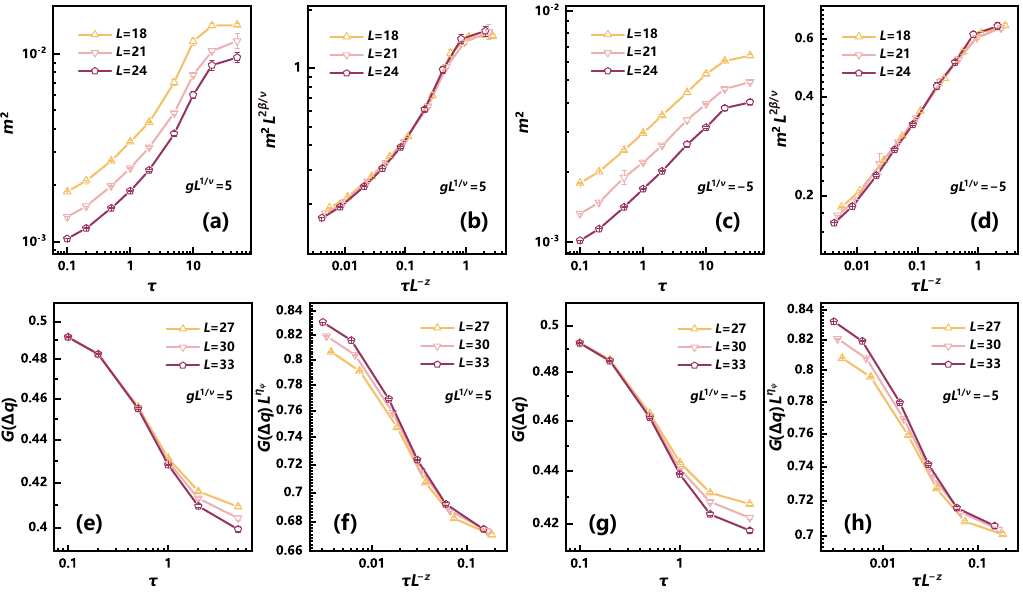}
        \caption{Relaxation dynamics of $m^2$ and $G$ from DSM initial state for $g\ne 0$.}
        \label{sfig:away_DSM}
    \end{figure}

    In addition, by fixing $g L^{1/\nu}$, we explore the iamgianry-time relaxation dynamics of $m^2$ and $G$ and show the results in Fig.~\ref{sfig:away_AFM}. For both $g L^{1/\nu}=5$ or $g L^{1/\nu}=-5$ (where $\nu=1.025$ estimated from Fig.~\ref{sfig:AFM_m2U}), we find that the rescaled curves of $m^2$ and $G$ according to Eqs.~(\ref{aaa}) and (\ref{bbb}), respectively, collapse onto each other quite well, confirming Eqs.~(\ref{aaa}) and (\ref{bbb}).
   
    Then we study the relaxation dynamics with DSM initial state. For both $g L^{1/\nu}=5$ or $g L^{1/\nu}=-5$ (where $\nu=1.025$), we find in Fig.~\ref{sfig:away_DSM} that the rescaled curves of $m^2$ collapse together according to Eq.~(\ref{ccc}). In addtion, for both $g L^{1/\nu}=5$ or $g L^{1/\nu}=-5$ with $\nu=1.025$, the rescaled curves of $G$ match with each other, confirming Eq.~(\ref{ddd}).

\section{Fermion correlation and quasi-particle weight}

    The quasi-particle weight $Z$ characterizes fermionic single-particle excitations near the Fermi surface/Dirac point~\cite{Sorella2019prb,Lang2019PRL,Vaezi2022PRL,Migdal1957JETP}. The relation between $Z$ and the distribution function is
    \begin{equation}
        Z = \lim_{L \to \infty} [n(\varepsilon_{\bm{q}} = 0^-) - n(\varepsilon_{\bm{q}} = 0^+)],
    \end{equation}
    in which the energy resolved momentum distribution function is defined as
    \begin{equation}
        n(\varepsilon_{\bm{q}}=\pm|\varepsilon_{\bm{q}}|)
        =\braket{\psi_{\pm,\bm{q},\sigma}^\dagger\psi_{\pm,\bm{q},\sigma}}.
    \end{equation}
    $\psi_{\pm,\bm{q},\sigma}$ is the quasi-particle operator, with $+(-)$ indicating above (below) the Fermi surface/Dirac point, respectively, and $\sigma$ being the spin index.

    For the non-interacting case, the quasi-particle energy takes the eigenvalues of the Hamiltonian $\varepsilon_{\bm{q}}=\pm|h_{\bm{q}}|$, where $h_{\bm{q}}=-t(1+\expup{-\i \q\cdot\bm{a}_1}+\expup{-\i \q\cdot\bm{a}_2})$ in honeycomb lattice, and the Hamiltonian can be written into diagonal representation:
    \begin{equation}
        H_0=\sum_{\bm{q},\sigma} 
        \mat{c_{A,\bm{q},\sigma}^\dagger&c_{B,\bm{q},\sigma}^\dagger}
        \mat{ 0 & h_{\bm{q}} \\ h^*_{\bm{q}} & 0 }
        \mat{c_{A,\bm{q},\sigma}\\c_{B,\bm{q},\sigma}}
        =\sum_{\bm{q},\sigma} 
        \mat{\psi_{+,\bm{q},\sigma}^\dagger&\psi_{-,\bm{q},\sigma}^\dagger}
        \mat{ |h_{\bm{q}}|&0 \\ 0&-|h_{\bm{q}}| }
        \mat{\psi_{+,\bm{q},\sigma}\\\psi_{-,\bm{q},\sigma}},
    \end{equation}
    where the annihilation operators $\psi_{-,\bm{q},\sigma}$ and $\psi_{+,\bm{q},\sigma}$ of the bonding and anti-bonding states, respectively, are given as
    \begin{equation}
        \psi_{\pm,\bm{q},\sigma}=\frac{1}{\sqrt{2}}\kuohao{c_{A,\bm{q},\sigma}\pm\frac{h_{\bm{q}}}{|h_{\bm{q}}|}c_{B,\bm{q},\sigma}}.
    \end{equation}
    The orbital subscripts $A, B$ represent the two sublattices in the honeycomb, and $c_{\alpha,\bm{q},\sigma}$ annihilation an electron on $\alpha=A,B$ sublattice:
    \begin{equation}
        c_{\alpha,\bm{q},\sigma} = \frac{1}{\sqrt{N}}\sum_{i_\alpha}\expup{-\i\q\cdot\bm{r}_i}c_{i_\alpha,\sigma}.
    \end{equation}
    Thus, the occupation in the non-interacting case simply reads:
    \begin{equation}
        \braket{\psi_{\pm,\bm{q},\sigma}^\dagger\psi_{\pm,\bm{q},\sigma}}
        =\frac{1}{2}\kuohao{\braket{c_{A,\bm{q},\sigma}^\dagger c_{A,\bm{q},\sigma}}+\braket{c_{B,\bm{q},\sigma}^\dagger c_{B,\bm{q},\sigma}}}
        \pm \frac{1}{2}\kuohao{\frac{h_{\bm{q}}}{|h_{\bm{q}}|}\braket{c_{A,\bm{q},\sigma}^\dagger c_{B,\bm{q},\sigma}}+\frac{h_{\bm{q}}^*}{|h_{\bm{q}}|}\braket{c_{B,\bm{q},\sigma}^\dagger c_{A,\bm{q},\sigma}}}.
    \end{equation}
    Considering the non-interacting solution $\braket{c_{A,\bm{q},\sigma}^\dagger c_{A,\bm{q},\sigma}}=\braket{c_{B,\bm{q},\sigma}^\dagger c_{B,\bm{q},\sigma}}=\frac{1}{2}$, $\braket{c_{A,\bm{q},\sigma}^\dagger c_{B,\bm{q},\sigma}}^*=\braket{c_{B,\bm{q},\sigma}^\dagger c_{A,\bm{q},\sigma}}=-\frac{h_{\bm{q}}}{2|h_{\bm{q}}|}$ at half filling, the groud-state-occupation of the bonding and anti-bonding are respectively $\braket{\psi_{-,\bm{q},\sigma}^\dagger\psi_{-,\bm{q},\sigma}}=1$ and $\braket{\psi_{+,\bm{q},\sigma}^\dagger\psi_{+,\bm{q},\sigma}}=0$. Consequently, quasi-particle weight $Z=1$. 

    For the case with interaction, $\braket{c_{A,\bm{q},\sigma}^\dagger c_{A,\bm{q},\sigma}}=\braket{c_{B,\bm{q},\sigma}^\dagger c_{B,\bm{q},\sigma}}=\frac{1}{2}$ for all momenta still holds due to the particle-hole symmetry at half-filling, 
    and only off-diagonal correlation $\braket{c_{A,\bm{q},\sigma}^\dagger c_{B,\bm{q},\sigma}}^*=\braket{c_{B,\bm{q},\sigma}^\dagger c_{A,\bm{q},\sigma}}\equiv f_{\q}$ bears non-trivial fermion correlation. Therefore, the "dressed" quasi-particle operators in interacting case read (the same formula appears in Ref. \cite{Sorella2016prx}):
    \begin{equation}
        \psi_{\pm,\bm{q},\sigma}=\frac{1}{\sqrt{2}}\kuohao{c_{A,\bm{q},\sigma}\pm\frac{f_{\bm{q}}}{|f_{\bm{q}}|}c_{B,\bm{q},\sigma}},
    \end{equation}
    with the occupation
    \begin{equation}
        \braket{\psi_{\pm,\bm{q},\sigma}^\dagger\psi_{\pm,\bm{q},\sigma}}=\frac{1}{2}\pm |f_{\bm{q}}|.
    \end{equation}
    Finally, one can use the fermion correlation near the Fermi surface (Dirac point $K$) to calculate the occupation jump (Ref. \cite{Lang2019PRL}):
    \begin{equation}
        Z = \lim_{L \to \infty} 2|f_{K+\Delta \bm{q}}| = \lim_{L \to \infty} 2|G(\Delta \bm{q})|
    \end{equation}
    where $G(\Delta \bm q) \equiv \frac{1}{L^d} \sum_{ij} \mathrm{e}^{\mathrm{i} (K+\Delta \bm {q}) \cdot (\bm{r}_i - \bm{r}_j)} c_{i,A}^\dagger c_{j,B}$ is the fermion correlation we calculated in the main text, and $\Delta \bm{q}$ is the smallest lattice momentum. Due to the direct proportional relationship, the quasiparticle weight $Z$ and the fermion correlation $G(\Delta \bm{q})$ have the same anomalous dimension $\eta_\psi$. In the thermodynamic limit, the DSM phase has $Z=1$, meaning $G(\Delta \bm{q}) = 0.5$, while the AFM phase has $Z=0$, with $G(\Delta \bm{q}) = 0$. 

\end{document}